\documentstyle[titlepage,12pt]{article}
\begin{document}
\newfont{\bg}{cmr10 scaled\magstep3}
%
\newcommand{\gsimeq}
{\hbox{ \raise3pt\hbox to 0pt{$>$}\raise-3pt\hbox{$\sim$} }}
\newcommand{\lsimeq}
{\hbox{ \raise3pt\hbox to 0pt{$<$}\raise-3pt\hbox{$\sim$} }}


\newcommand{\plb}[3]{Phys. Lett. {\bf B#1}, #3, (#2)} 
\newcommand{\prl}[3]{Phys. Rev. Lett. {\bf #1}, #3, (#2)}
\newcommand{\prd}[3]{Phys. Rev. {\bf D#1}, #3, (#2)}
\newcommand{\npb}[3]{Nucl. Phys. {\bf B#1}, #3, (#2)}
\newcommand{\npbps}[3]{Nucl. Phys. {\bf B}(Proc. Suppl.) {\bf #1}, #3, (#2)}
\newcommand{\prog}[3]{Prog. Theor. Phys. {\bf #1}, #3, (#2)}
\newcommand{\zeitc}[3]{Z. Phys. {\bf C#1}, #3, (#2)}
\newcommand{\mpl}[3]{Modern. Phys. Lett. {\bf A#1}, #3, (#2)} 
\newcommand{\ijmp}[3]{Int. J. Modern. Phys. Lett. {\bf A#1}, #3, (#2)}


\newcommand{\psibar}{\bar{\psi}}
\newcommand{\upp}{u_{+}(p,\sigma)} 
\newcommand{\vmp}{v_{-}(p,\sigma)} 
\newcommand{\umbq}{\bar{u}_{-}(q,\tau)} 
\newcommand{\vpbq}{\bar{v}_{+}(q,\tau)} 
\newcommand{\umbp}{\bar{u}_{-}(p,\sigma)} 
\newcommand{\vpbp}{\bar{v}_{+}(p,\sigma)} 
\newcommand{\upk}{u_{+}(k,s)} 
\newcommand{\vmk}{v_{-}(k,s)} 
\newcommand{\umbk}{\bar{u}_{-}(k,s)} 
\newcommand{\vpbk}{\bar{v}_{+}(k,s)} 
\newcommand{\upmk}{u_{\pm}(k,s)} 
\newcommand{\vpmk}{v_{\pm}(k,s)} 
\newcommand{\upmbk}{\bar{u}_{\pm}(k,s)} 
\newcommand{\vpmbk}{\bar{v}_{\pm}(k,s)} 
\newcommand{\bp}{b^\dagger_{+}(q,\tau)}
\newcommand{\bm}{b_{-}(p,\sigma)}
\newcommand{\ddp}{d^\dagger_{+}(p,\sigma)}
\newcommand{\dm}{d_{-}(q,\tau)}
\newcommand{\cbp}{\cos{\beta(p)}}
\newcommand{\cbq}{\cos{\beta(q)}}
\newcommand{\cbk}{\cos{\beta(k)}}
\newcommand{\gfive}{\gamma_5}
\newcommand{\gfivetc}{\gamma_5 T_c}  
\newcommand{\no}{{\bf:}\Omega{\bf:}}
\newcommand{\nopq}{{\bf:}\Omega(p,q){\bf:}}
\newcommand{\vone}{{\cal V}_1}
\newcommand{\vtwo}{{\cal V}_2}
\newcommand{\gp}{G_{+}}
\newcommand{\gm}{G_{-}}
\newcommand{\ps}{\langle + \vert}
\newcommand{\ms}{\vert - \rangle }
\newcommand{\splus}{S_{+}(p)}
\newcommand{\sm}{S_{-}(p)}
\newcommand{\notp}{\!\!\not\!p}
\newcommand{\calv}{{\cal V}}
\newcommand{\calh}{{\cal H}} 
\newcommand{\vmu}{V_{1\mu}(p+k)}
\newcommand{\vnu}{V_{1\nu}(p+k)}
\newcommand{\dmunu}{D_{\mu\nu}(p-k)}
\newcommand{\wplus}{\omega_{+}(p)+ \omega_{+}(k)} 
\newcommand{\wminus}{\omega_{-}(p)+ \omega_{-}(k)} 
\newcommand{\wpm}{\omega_{\pm}(p)+ \omega_{\pm}(k)} 
\newcommand{\wtwo}{2\omega_{\pm}(p)} 
\newcommand{\gb}{\bar{g}^2}
\newcommand{\svp}{\vert + \rangle }
\newcommand{\svm}{\vert-\rangle }
\newcommand{\svpm}{\vert\pm\rangle }
\newcommand{\svap}{\vert A+\rangle}
\newcommand{\svam}{\vert A-\rangle}
\newcommand{\svapm}{\vert A \pm\rangle}
\newcommand{\svpd}{\langle +\vert}
\newcommand{\svmd}{\langle -\vert}
\newcommand{\sapd}{\langle A+\vert}
\newcommand{\samd}{\langle A-\vert}
\newcommand{\svpmd}{\langle \pm \vert}
\newcommand{\svapmd}{\langle A \pm \vert}

\begin{titlepage}
\rightline{}
\rightline{}
\rightline{}

\vspace*{2cm}

\addtocounter{footnote}{1}

\begin{center}

{\large \bf 

Renormalizability of a lattice chiral fermion  \\ 

\vspace{0.5cm}
in the overlap formulation
 }
\\
\vspace{0.5cm}

\vspace{2cm}
{\sc Atsushi Yamada}\footnote{On leave of absence from: 
Department of Physics, University of Tokyo, Tokyo, 113 Japan. }  
\\
\vspace{1cm}
{\it International Center for Theoretical Physics, \\
\vspace{0.5cm}
     Miramare, Trieste, Italy  }
\vspace{2cm}

{\bf ABSTRACT}
\end{center}

Renormalizability of a lattice chiral fermion 
is studied at one loop level in the overlap formulation in four dimensions. 
The fermion chirality is examined including the self-energy corrections 
due to gauge interactions. 
Divergent terms breaking the chiral symmetry 
do not appear and the chiral fermion 
is renormalized, preserving the correct chiral properties without 
adding new counter-terms or tuning the parameters involved. 
The divergent part of the wave function 
renormalization factor agrees with that of the continuum theory. 
The overlap formulation of a lattice chiral fermion has passed   
the important test, the renormalizability, at one loop level. 

\end{titlepage}

\baselineskip = 0.7cm

Kaplan's original idea of the domain wall fermion \cite{kaplan} 
inspired many studies of lattice 
formulations of a chiral fermion \cite{chiral,over}. 
The overlap formulation \cite{over} 
has been shown to have promising analytic properties 
\cite{aoki}-\cite{kikukawa} 
and gives results in good agreement with continuum theories 
in the numerical simulations of two dimensional systems \cite{num}. 
This formulation provides the possibility of investigating 
chiral gauge theories 
in the strong coupling region, as well as a clearer analysis of the 
phenomenon of chiral symmetry breaking 
in vector gauge theories like QCD. Moreover, 
it has been  recently argued that this formulation can be applied in the 
study of supersymmetric gauge theories \cite{susy} and 
strongly correlated fermion systems in three (two plus one) 
dimensions \cite{3d}. 
However, one of the most important tests of the validity of that formulation, 
the renormalizability of a chiral fermion, has not yet been examined. 
Since the existence of the triangle
anomaly means that it is impossible to regularize a chiral fermion  
in a chiral invariant way \cite{smit}, 
even if a chiral fermion is obtained in the continuum limit 
in a (lattice) regularization at tree level \cite{nn}, 
it is not evident that the chirality of the regularized fermion is 
preserved after including quantum corrections. 
The fact that the chiral anomaly is correctly reproduced 
in the overlap formulation \cite{anomaly} shows that the chirality of the 
regularized fermion is preserved in the triangle anomaly diagram, where 
the dynamics of gauge fields does not play any role. 
However, this fact does not guarantee that the chirality 
of the regularized fermion is preserved after including 
the radiative corrections by the dynamics of the gauge bosons \cite{wilson}. 
In this letter, we study the fermion propagator 
in the overlap formulation regularized on a lattice, 
including self-energy corrections due to the $SU(N)$ gauge interactions 
at one loop level, and directly examine the chirality of the fermion and its
renormalizability.


In the overlap formulation, the effective action of a chiral fermion in 
the presence of gauge fields 
is expressed by the overlap of the two vacua  $\svapm$ of the Hamiltonians 
${\cal H}_{\pm}(A)$, where  
\begin{eqnarray}
& &{\cal H}_{\pm}(A) = \int_{p} \psi^\dagger (p) H_{\pm}(p) \psi(p) +
{\cal V}(A), 
\label{eqn:H}
\\
& &H_{\pm}(p)=\gfive\Bigl[\sum_{\mu=1}^{4} i {\tilde p}_\mu \gamma_\mu + T_c
X_{\pm}(p) \Bigr] , \,\,\,\,\,\,
X_{\pm}(p)= \pm\frac{\lambda}{a} + \frac{ar}{2} \hat{p}^2.
\label{eqn:h}
\end{eqnarray}
In eq. (\ref{eqn:H}), the momentum integral is 
over the Brillouin zone $[-\pi/a,\pi/a]$ and the term ${\cal V}(A)$
describes the 
gauge interactions, which we treat as perturbations. 
In eq. (\ref{eqn:h}), 
$\tilde{p}_\mu=(1/a)\sin (p_\mu a)$, $\hat{p}_\mu=(2/a)\sin(p_\mu a/2)$, 
$a$ is the lattice spacing, and 
$T_c=\pm1$ determines the fermion chirality, as will be seen later.  
The Hamiltonians ${\cal H}_{\pm}$ describe time evolution of a 
Dirac fermion in four plus one dimensions and 
this Dirac fermion is reduced to a Weyl fermion 
in four dimensions \cite{over,n}. 
The operator $\psi(p)$ is expanded in terms of 
creation and annihilation operators as, 
\begin{eqnarray}
 \psi(p)=  \sum_{\sigma} \Bigl[ u_{\pm}(p,\sigma) b_{\pm}(p,\sigma) 
 + v_{\pm}(p,\sigma) d^\dagger_{\pm}(p,\sigma)  \Bigr],
\label{eqn:psi}
\end{eqnarray}
where $u_{\pm}(p,\sigma)$ and $v_{\pm}(p,\sigma) $ are positive and 
negative energy eigenspinors of the one-particle Hamiltonian $H_{\pm}(p)$, 
respectively, i.e., 
$H_{\pm}(p) u_{\pm}(p,\sigma) = \omega_\pm (p) u_{\pm}(p,\sigma)$ and 
$H_{\pm}(p) v_{\pm}(p,\sigma) = -\omega_\pm (p) v_{\pm}(p,\sigma)$. 
The label $\sigma$ denotes the spin states and 
$\omega_\pm (p) =\sqrt{{\tilde p}^2+ X^2_{\pm}(p) } $. 
The spinors $u_{\pm}$ and $v_{\pm}$ are given by
\begin{eqnarray}
u_{\pm}(p,\sigma) = 
\frac{\omega_{\pm} + X_{\pm}  -i \sum_{\mu}{\tilde  p}_\mu \gamma_\mu T_c}
{\sqrt{2\omega_{\pm}(\omega_{\pm} + X_{\pm})}} \chi(\sigma), 
\nonumber \\
v_{\pm}(p,\sigma) = 
\frac{\omega_{\pm} - X_{\pm}  + i \sum_{\mu}{\tilde  p}_\mu \gamma_\mu T_c}
{\sqrt{2\omega_{\pm}(\omega_{\pm} - X_{\pm})}} \chi(\sigma), 
\label{eqn:uv}
\end{eqnarray}
and the spinor $\chi(\sigma)$ satisfies $\gfivetc \chi(\sigma)=\chi(\sigma)$. 
The two sets ($b_{+}(p,\sigma)$, $d_{+}(p,\sigma)$) and 
($b_{-}(p,\sigma)$, $d_{-}(p,\sigma)$) are related by a Bogoliubov 
transformation as,  
\begin{eqnarray}
& &b_{-}(p,\sigma)= \cbp b_{+}(p,\sigma) -\sin\beta(p)
d^\dagger_{+}(p,\sigma), 
\nonumber \\
& &d^\dagger_{-}(p,\sigma)= \sin\beta(p) b_{+}(p,\sigma)+ \cbp 
d^\dagger_{+}(p,\sigma), 
\label{eqn:bog}
\end{eqnarray}
with $\cbp = u^\dagger_{+}(p,\sigma) u_{-}(p,\sigma) $. 
The Dirac vacua $\svpm$ of the Hamiltonian $\calh_{\pm}(0)$ 
are defined as $b_{\pm}(p,\sigma),d_{\pm}(p,\sigma) \svpm =0$, and 
the Dirac vacua $\vert A\pm\rangle$ are given by the integral equations, 
\begin{eqnarray}
\svapm &=& \alpha_\pm(A)\Bigl[1-G_\pm ({\cal V}-\Delta E_\pm)\Bigr]^{-1} \svpm,
\,\,\,\,\,\, 
G_\pm = \frac{1-\svpm \svpmd}{E_\pm(0)-H_\pm(0)},
\label{eqn:A}
\\
& &\vert \alpha_\pm(A)   \vert^2= 1- 
\svapmd \Bigl[{\cal V}-\Delta E_\pm\Bigr]G^2_\pm 
\Bigl[{\cal V}-\Delta E_\pm\Bigr] \svapm,
\label{eqn:al}
\end{eqnarray}
where $\Delta E_\pm =E_\pm (A) - E_\pm(0)$. $E_\pm (A)$ and  
$E_\pm(0) $ are the ground state energies of the Hamiltonians 
$\calh_\pm(A)$ and $\calh_\pm(0)$, respectively. 
The fermion propagator is defined by the path integral 
\begin{eqnarray}
\frac{\int {\cal D}{\cal A}  
\langle A+ \vert \Omega(p,q)\vert A- \rangle e^{-S(A)} }
{ \int {\cal D}{\cal A}    \langle A+ \vert A- \rangle e^{-S(A)} },
\label{eqn:path}
\end{eqnarray}
where 
$\Omega(p,q) =
\{\psi(p) \psibar(q) - \psibar(q)\psi(p)\}/2$, 
$\psibar=\psi^\dagger\gfive$ and $S(A)$ is 
the action of the gauge field. 
We find it convenient for later calculations 
to decompose $\Omega(p,q)$ in the following form (using 
the Bogoliubov transformation (\ref{eqn:bog})), 
$\Omega(p,q) = \svpd \Omega(p,q) \svm + \nopq $, 
\begin{eqnarray}
\svpd \Omega(p,q) \svm &=&  (2\pi)^4 \delta^{4}_P(p-q) 
\frac{1}{2}\Bigl[ S_{+}(p) -S_{-}(p)  \Bigr],
\label{eqn:tree}
\\
\nopq&=&\frac{1}{\cbp\cbq} \cdot
\sum_{\sigma,\tau} \Bigl[ - \upp\umbq \bp\bm  
\nonumber \\
& &+  \upp \vpbq \bm\dm  + \vmp\umbq\ddp\bp 
\nonumber \\
& &+ \vmp\vpbq\ddp\dm \Bigr].
\label{eqn:normal}
\end{eqnarray}
Here $\delta^{4}_P(p-q)$ is the periodic 
$\delta$-function on the lattice, 
$S_{+}(p)= \sum_{\sigma}\upp\umbp/\cbp$ and 
$S_{-}(p)= \sum_{\sigma}\vmp\vpbp/\cbp$. 
The first term (\ref{eqn:tree}) is the propagator for $A=0$ 
and the second term (\ref{eqn:normal}) 
satisfies the relation  $\svpd\nopq \svm =0$. 
Near the origin $p\simeq 0$, 
\begin{eqnarray}
S_{\pm}(p) \simeq \frac{\lambda}{a}\frac{1}{p^2}
\Bigr[ \frac{1}{2}(1+\gfivetc)(\mp i \notp) +\frac{a}{2\lambda}p^2 \gfive 
\Bigl] ,
\label{eqn:propagator}
\end{eqnarray}
and thus the propagator $[ S_{+}(p) -S_{-}(p)]/2 $ describes a chiral fermion 
in this region. 
At each corner of the Brillouin zone, $p_\mu \simeq  \pm \pi/a + q_\mu$, 
the propagator takes the following form,
\begin{eqnarray}
\frac{1}{2}\Bigl[ S_{+}(p) -S_{-}(p)  \Bigr]
\simeq \frac{1}{ \sqrt{(4 n^2 r^2 - \lambda^2)} + {\cal O}(a^2q^2) } 
(c_1+c_2\gfive)
\label{eqn:propagatorp}
\end{eqnarray}
where $n=1,\cdots,4$ 
is the number of momentum components which lie near the corner of the 
Brillouin zone and $c_{1,2}$ are constants. 
The gauge symmetry of the Hamiltonians (\ref{eqn:H}) 
becomes a chiral gauge symmetry near the origin of the 
Brillouin zone, while chiral non-invariant 
contributions coming from each corner are suppressed 
due to the $\sqrt{ (4 n^2 r^2-\lambda^2)}$ mass. 
(We restrict ourselves to the range of parameters $0< \lambda <2r$.)


Now we consider the one loop correction to the propagator 
$\svpd \Omega(p,q) \svm $. 
Inserting the decomposition of $\Omega(p,q)$ into eq. (\ref{eqn:path}), 
eq. (\ref{eqn:tree}) yields the propagator at tree level, 
while the quantum corrections arise from 
$\langle A+ \vert \nopq \vert A- \rangle$. 
To obtain the one loop correction, the interaction $\calv(A)$ in eq.
(\ref{eqn:H}) should be 
expanded up to the second order in the gauge coupling constant $g$;  
$\calv(A)=\calv_{1}(A) + \calv_{2}(A)$, where 
\begin{eqnarray}
& &{\cal V}_{1} = i g \int_{p,q} \psibar(p) \sum_{\mu} 
V_{1\mu}(p+q) A_\mu(p-q) \psi(q),
\label{eqn:int1}
\\
& &{\cal V}_2 = \frac{1}{4} a g^2 \int_{p,t,q} \psibar(p)\sum_{\mu,\nu} 
V_{2\mu}(p+q)\delta_{\mu\nu}\{A_{\mu}(t),A_{\nu}(p-t-q)\}\psi(q), 
\label{eqn:int2}
\end{eqnarray}
$V_{1\mu}(p) = \gamma_{\mu} \cos(p_\mu a/2) -ir \sin(p_\mu a/2)$, 
and $V_{2\mu}(p)=  r\cos(p_\mu a/2) -i\gamma_\mu \sin(p_\mu a/2) $. 
Then evaluating $\svapm$ perturbatively in eq. (\ref{eqn:A}) the quantum
correction 
$\langle A+ \vert \no \vert A- \rangle$,
up to the order $g^2$ \cite{alpha}, is 
\begin{eqnarray}
& & \ps \no \gm \vtwo \ms +  \ps \vtwo \gp \no \ms + 
 \ps \vone \gp \no \gm \vone \ms 
\nonumber \\
& &+ \ps \vone \gp \vone \gp \no \ms + 
\ps \no \gm \vone \gm \vone \ms .  
\label{eqn:tot}
\end{eqnarray}
These terms are evaluated by rewriting the fermion operators in 
$\vone$ and $\vtwo$ in terms of 
the creation and annihilation operators defined in eq. (\ref{eqn:psi}). 
Then performing the path integral over the gauge fields, 
the first two terms and the last three terms lead to the 
quantum corrections described by the 
Feynman diagrams Figs. 1. (a) and (b), respectively. 

Before examining these terms separately, we briefly 
discuss their general features and the strategy of our analysis. 
Each term in eq. (\ref{eqn:tot}) yields the contribution to the 
tree level propagator of the form 
$S_{\varepsilon}(p) a \Sigma(p) S_{\varepsilon'}(p)$ where 
$\varepsilon$ and $\varepsilon'$ denote $\pm$, and 
$ \Sigma(p) $ is a self-energy given by integral over the loop momentum. 
Here, we explicitly factor out the lattice spacing $a$ so that 
$\Sigma(p) $ has the correct dimension (one) of the self-energy for 
fermions. 
To compute the divergent part of the renormalization factors, 
$ \Sigma(p)$ should be evaluated up to the logarithmically 
divergent part in the continuum limit. 
We find that in general, in this limit 
$\Sigma(p)$ gives rise to the 
following expression, 
\begin{eqnarray}
\Sigma(p) &=& \frac{1}{a} \Bigl[
\sigma_{1}(\lambda,r,\xi) + \sigma_{2}(\lambda,r,\xi)
\gfive \Bigr] 
+ \frac{1}{\lambda}C \frac{\gb}{16\pi^2} \{1-(1-\xi)\}\log(p^2a^2)\frac{1}{2}
(1-\gfivetc)i\notp
\nonumber \\
& &+ (finite\,\,\, terms), 
\label{eqn:limit}
\end{eqnarray}
where $\sigma_{1,2}(\lambda,r,\xi)$ are the constants obtained 
by the integration over the loop momentum 
and $C$ is a numerical coefficient. 
Here $\xi$ is the gauge fixing parameter, defined later, 
and $\gb$=$(N^2-1)/(2N) g^2 $. 
This expression is obtained, for example, 
by expanding $\Sigma(p) $ with respect to 
the external momentum $p$ as in Refs. \cite{smit,kawai,kawamoto}, or by 
splitting the integration region of the loop momentum 
into two pieces as in Ref. \cite{smit}. 
In both methods, the ultraviolet divergent parts are evaluated 
by explicitly factoring out the $a$ dependence from the momentum 
integral with a rescaling of the loop momentum 
$k \rightarrow \bar{k}/a$. Logarithmic 
divergences appear as the infrared divergences 
of the integration with respect to the rescaled variable $\bar{k}$. 
Following the strategy of Refs. \cite{smit,kawai,kawamoto}, we expand 
$\Sigma(p) $ up to the first order in $p$, where zero 
and first order terms give rise to the ultraviolet divergence 
in the continuum limit as,    
\begin{eqnarray}
\Sigma(0) &\rightarrow& \frac{1}{a} \Bigl[
\sigma_{1}(\lambda,r,\xi) + \sigma_{2}(\lambda,r,\xi)
\gfive \Bigr]  + finite\,\,\,term,
\label{eqn:limita}
\\
\sum_{\mu} p_\mu 
\frac{\partial \Sigma(p)}{\partial p_\mu} \Bigl|_{p =0} 
&\rightarrow& 
\frac{1}{\lambda}C \frac{\gb}{16\pi^2} \{1-(1-\xi)\}\log(\kappa^2a^2)\frac{1}{2}
(1-\gfivetc)i\notp + finite\,\,\, term, 
\label{eqn:limitb}
\end{eqnarray}
where $\kappa$ is the infrared regulator. 
The remaining finite part is given by 
\begin{eqnarray}
& &\Sigma(p) - \Sigma(0) - \sum_{\mu} p_\mu 
\frac{\partial \Sigma(p)}{\partial p_\mu} \Bigr|_{p =0} 
\nonumber \\  
\rightarrow 
& &
\frac{1}{\lambda}C \frac{\gb}{16\pi^2} \{1-(1-\xi)\}\log(p^2/\kappa^2)\frac{1}{2}
(1-\gfivetc)i\notp.
\label{eqn:limitc}
\end{eqnarray}
Expression (\ref{eqn:limit}) is the sum of eqs. (\ref{eqn:limita}), 
(\ref{eqn:limitb}) and (\ref{eqn:limitc}). 
The infrared divergence at $\kappa \rightarrow 0$ is cancelled 
between eqs. (\ref{eqn:limitb}) and (\ref{eqn:limitc}). 
In eq. (\ref{eqn:limitb}), 
the logarithmic divergence is obtained if the integral over the 
rescaled loop momentum $\bar{k}=ak$ exhibits infrared divergences. 
When the integral over $\bar{k}$ is infrared finite, there is no logarithmic 
term in eq. (\ref{eqn:limitc}). 
The renormalization factors are evaluated by inserting 
eq. (\ref{eqn:limit}) into the expression 
$S_{\varepsilon}(p) a \Sigma(p) S_{\varepsilon'}(p)$. 
In eq. (\ref{eqn:limit}), 
the Lorentz structure of the linearly divergent terms is 
completely determined by the discrete Lorentz symmetry on the lattice 
e.g., ($k \leftrightarrow -k $). 
These terms are reduced to a 
finite wave function renormalization factor, 
\begin{eqnarray}
& &
\frac{\lambda}{a}\frac{1}{p^2}
\Bigl[ \frac{1}{2}(1+\gfivetc)(-\varepsilon i \notp) 
+\frac{a}{2\lambda}p^2 \gfive 
\Bigr] a \frac{1}{a} \Bigl[     
\sigma_{1} + \sigma_{2} \gfive \Bigr]   
\cdot 
\frac{\lambda}{a}\frac{1}{p^2}\Bigl[\frac{1}{2}(1+\gfivetc)(-\varepsilon'
i \notp) + 
 \frac{a}{2\lambda}p^2 \gfive \Bigr] 
\nonumber \\ 
\rightarrow & &
\frac{1}{2}\Bigl[ \sigma_1 T_c (-\varepsilon + \varepsilon'   )
+ \sigma_2 (\varepsilon + \varepsilon'   ) +{\cal O}(a) \Bigr]
\frac{\lambda}{a}\frac{1}{2}(1+\gfivetc)\frac{-i \notp}{p^2}. 
\label{eqn:wavefini}
\end{eqnarray}
From this expression, we can see that only an ultraviolet divergence of 
$\Sigma(p) $ more severe than quadratic can invalidate the 
chiral property of the regularized fermion. 
The logarithmically divergent term in eq. (\ref{eqn:limitb}) directly 
appears in the wave function renormalization factor as 
a logarithmic divergence.


Next we discuss each contribution in eq. (\ref{eqn:tot}) separately.  
The contributions of the first two terms to the tree level propagator 
are given by $-S_{\mp}(p) a \Pi^a_{\pm}(p) S_{\pm} (p) $ with 
\begin{eqnarray}
\Pi^a_{\pm}(p) = -\frac{\gb}{2} \frac{1}{\wtwo} \sum_{\mu} 
V_{2 \mu}(2p) \int_{k} D_{\mu\mu}(k) 
\rightarrow  \frac{\gb}{4\lambda} \sigma^a (\xi) (\frac{4r}{a}-i \notp),
\label{eqn:pia}
\end{eqnarray}
where the upper (lower) sign should be taken for 
the contribution containing $G_+$ ($G_{-}$), and 
$D_{\mu\nu}(k)$ is the gauge boson propagator,
\begin{eqnarray}
D_{\mu\nu}(k)= \frac{1}{\hat{k}^2}
(\delta_{\mu\nu} - (1-\xi)\frac{\hat{k}_\mu\hat{k}_\nu }{\hat{k}^2}).
\end{eqnarray}
In eqs. (\ref{eqn:pia}), 
$\sigma^a(\xi)$ is the constant obtained by the integration over
the rescaled loop momentum $\bar{k}=ak$. No logarithmic divergence
appears, since the rescaled integral is infrared finite. 
As was discussed previously, the contributions (\ref{eqn:pia}) 
are reduced to a finite wave function renormalization factor. 

Next consider the last three terms in eq. (\ref{eqn:tot}). 
The third term leads to the following contributions to the tree level 
propagator; 
$\splus a \Sigma^{b}_{+}(p) \splus + \sm a \Sigma^{b}_{-}(p)\sm $
with 
\begin{eqnarray}
\Sigma^{b}_{\pm}(p) =  
\pm  \frac{1}{a} \gb \sum_{\mu,\nu} \int_{k} \Bigl[ \frac{1}{\wplus} \Bigr]
\Bigl[ \frac{1}{\wminus}\Bigr]
\vmu S_{\mp}(k) \vnu \dmunu.
\label{eqn:sigmaa}
\end{eqnarray} 
The integration over the rescaled momentum $\bar{k}$ exhibits 
infrared divergences near $\bar{k}\simeq0$, which in turn, give rise 
to logarithmic divergences in the form of eq. (\ref{eqn:limit}) with 
$C=1/4$. Infrared divergences 
appear only in the region $\bar{k}\simeq0$, as 
is seen from eqs. (\ref{eqn:propagator}) and 
(\ref{eqn:propagatorp}). 
The contributions of the fourth and fifth terms in eq. (\ref{eqn:tot})
are expressed as  
$- S_{\mp}(p) a [\Sigma^{b'}_{\pm}(p) + \Pi^{b}_{\pm}(p) ] S_{\pm}$, 
where 
\begin{eqnarray}
\Sigma^{b'}_{\pm}(p) &=& \frac{1}{a} \gb \sum_{\mu,\nu} \int_{k} 
\Bigl[ \frac{1}{\wpm} \Bigr]^2
\vmu \frac{\sin\beta(k)}{\cbk}\sum_{s} \Bigl[
\begin{array}{r} 
\upk\vpbk \,\,\, (for \,\,\,+) 
\\
-\vmk \umbk  \,\,\, (for \,\,\,-)
\end{array} 
\Bigr] 
\nonumber \\
& &\cdot \vnu \dmunu,
\label{eqn:sigmab}
\\
\Pi^{b}_{\pm}(p) &=& - \frac{1}{a} \gb \frac{1}{\wtwo} 
\sum_{\mu,\nu}
\int_{k} 
\frac{1}{\wpm}
\vmu 
\nonumber
\\
& &\cdot \sum_{s} \Bigl[ \upmk\upmbk - \vpmk \vpmbk \Bigr] \vnu \dmunu.
\label{eqn:pib}
\end{eqnarray}
The upper (lower) sign should be taken for
the contribution containing $G_+$ ($G_{-}$). 
The contributions (\ref{eqn:sigmab}) lead to linearly divergent 
terms as well as logarithmically divergent terms with $C=1/4$ 
in the form of eq. (\ref{eqn:limit}), while 
the contributions (\ref{eqn:pib}) lead only to the linearly divergent 
terms. This is because the integrations over the rescaled loop 
momentum $\bar{k}$ 
give rise to infrared divergences in eqs. (\ref{eqn:sigmab}), whereas 
in eqs. (\ref{eqn:pib}) they are finite.

Summing up all the contributions 
$\Pi^{a,b}_{\pm}$, $\Sigma^{b,b'}_{\pm}$ together with the tree level 
propagator, the propagator is given at one loop level as, 
\begin{eqnarray}
\frac{\lambda}{a}\frac{1}{2}(1+\gfivetc)\frac{-i\notp}{p^2} 
\Bigl[ 1 + \frac{\gb}{16\pi^2}\{1-(1-\xi)\}( \log a^2p^2 + finite
\,\,\,terms)\Bigr].  
\label{eqn:pone}
\end{eqnarray}
From this expression, we see that the chirality of the regularized fermion 
is properly preserved and the wave function renormalization factor is 
$Z = 1+ \frac{\gb}{16\pi^2}\Bigl[1-(1-\xi) \Bigr] ( \log a^2\mu^2 + const))$, 
where $\mu$ is the renormalization scale. 
The divergent part of the wave function renormalization factor agrees 
with that of the continuum theory.


We have studied the renormalization of a lattice chiral fermion due to the 
non-Abelian gauge interactions in the overlap formulation in four dimensions. 
Divergent terms breaking the chiral symmetry do not appear 
at one loop level, and 
accordingly there is no need to add new counter-terms or 
to tune parameters in the theory to specific 
values to realize a chiral fermion. The divergent part of the wave function 
renormalization factor is correctly reproduced. 
Our study proves the renormalizability of a lattice chiral fermion 
in the overlap formulation at one loop level, and 
indicates, together with the analyses of 
the gauge boson $n$-point functions \cite{vac,n} the renormalizability of this 
formulation. 


We would like to thank S.~Randjbar-Daemi, Y.~Kikukawa and R.~Narayanan 
for discussions and useful comments on the manuscript, 
and G.~Gyuk, M.~O'Loughlin and A.~Rasin for 
useful comments on the manuscript.


\section*{Figure Caption}
\renewcommand{\labelenumi}{Fig.~\arabic{enumi}}
\begin{enumerate}
\item The Feynman diagrams describing the contributions 
of (a) the first two terms 
and (b) the last three terms in eq. (\ref{eqn:tot}). 
\end{enumerate}
\section*{Figures}
\input FEYNMAN
\begin{picture}(8000,8000)
\drawline\fermion[\E\REG](0,0)[5000]
\drawloop\gluon[\W 8](\pbackx,\pbacky)
\drawline\fermion[\E\REG](\pbackx,\pbacky)[4000]
\drawline\fermion[\W\REG](\pbackx,\pbacky)[7000]
\end{picture}

\vspace{.5cm}

Fig. 1. (a).

\vspace{2cm}

\begin{picture}(8000,8000)
\drawline\fermion[\E\REG](0,0)[2000]
\drawloop\gluon[\NE 3](\pbackx,\pbacky)
\drawline\fermion[\E\REG](\pbackx,\pbacky)[2000]
\drawline\fermion[\W\REG](\pbackx,\pbacky)[7000]
\end{picture}

\vspace{.5cm}

Fig. 1. (b).


\begin{thebibliography}{99}
\bibitem{alpha} 
The functionals 
$\alpha_{\pm}(A) $ of $A$ in eq. (\ref{eqn:al}) are expanded as, 
$\alpha_{\pm}(A) \simeq 1 + {\cal O}(g^2 A^2)$ including 
complex phases and do not contribute to 
$langle A+ \vert \nopq \vert A- \rangle$ in this order.
\end{thebibliography}
\end{document}